\documentclass[12pt]{article}
\usepackage[T2A]{fontenc}
\usepackage[utf8]{inputenc}
\usepackage{amssymb,amsfonts,amsmath,mathtext,enumerate,float}
\usepackage[dvipdfm]{graphicx}
\usepackage{url}
\usepackage{authblk}
\usepackage{indentfirst}
\usepackage{pdfpages}
\date{}
\begin{document}

\begin{center}
%\title
{\hbox{\normalsize UDK 523.92+51.71}\large The Structure of the Solar Cycle and of the Activity Cycles of Late-Type Stars}

V.N.~Obridko$^1$, D.D.~Sokoloff$^{1,2}$, M.M.Katsova$^3$

$^1$IZMIRAN, Troitsk, Moscow, Russia\\
$^2$Lomonosov Moscow State University, Moscow, Russia\\
$^3$Sternberg Astronomical Institute of the Lomonosov Moscow State University, Moscow, Russiaя

\end{center}
%\maketitle

\centerline{Abstract}

It is shown that the description of the solar cycle that takes into account the odd zonal harmonic of the solar magnetic field allows us to deepen our knowledge of two important aspects of the solar activity. First, to clarify and expand predictions of the evolution of the cyclic activity of the Sun in the near future. Second, to develop a program for monitoring the spectrophotometric characteristics of radiation of the solar-type stars aimed at obtaining new information about their magnetic fields. 


\section*{Introduction}

The classical description states that the solar activity cycle is the interaction between the main dipole component of the solar poloidal magnetic field and the toroidal magnetic field. Here, the physical nature of these components is assumed to be intuitively clear and is not subjected to special analysis. However, the real extended solar cycle is much more complicated than the models and requires a detailed description. In this work, we develop the concept of an extended solar cycle by finding out what zonal harmonics are responsible for the features of the equatorial and polar propagation in the activity tracers on the surface. As a result, we arrive at  the conclusion that the zonal harmonics with $l = 5$ play a decisive role in separating the phenomena of both types associated with odd zonal harmonics.

According to the original scheme, the maximum of the equatorial field coincides with the inversion of the polar field, and both hemispheres are antisymmetric in the field sign. However, this simple and clear concept of the mean field does not fully explain a number of features of the solar activity. In particular, why are there cycles of different heights, how do spots arise from the mean field, and what is the nature of active longitudes?  This scheme and its generalizations deal with zonal characteristics; therefore, supersynoptic maps are analyzed, on which any dependence on longitude disappears.

In our research, we applied this description to the concept of an extended solar cycle. This concept was first formulated in 1988 [1, 2]. Shortly before this, observations appeared showing that the magnetic activity of one cycle could coexist with the magnetic activity of the previous cycle, and the overlapping interval could be up to seven years [3]. As seen from observations of the solar surface, the extended solar cycle begins at high latitudes during the sunspot maximum and consists of a relatively short poleward branch called ''Rush-to-the-Poles'' (RTTP) and a long branch directed towards the equator and passing through the solar minimum and the following solar cycle ([4], see also [5] and references therein).

The idea of an extended solar cycle as a single entity formed by the observed space-time structures was the result of many years of observations and a hard work on compiling catalogues carried out by the generation of outstanding observers of the past.

Some of these structures, such as prominences and filaments (e.g., see [6-9]), ephemeral active regions [10], and most features observed in the solar coronal green line [11], are naturally interpreted as magnetic phenomena. The origin of the other, such as the zonal structures of the torsion oscillations [12-14], is not yet completely clear. We should also mention long-term series of geomagnetic data, in which features are found that are consistent with the pattern of solar magnetic cycles overlapping in time [15, 16].

 In the recent work of our team [17], we made the following additions to the concept of the extended solar cycle:
 
1. The initial paradigm of the solar dynamo is reasonably consistent with the description of the 11-year cycle of solar activity using the first two odd zonal harmonics of the large-scale solar magnetic field.

 2. On the other hand, this description needs a modification and refinement within the concept of the extended solar cycle. During the overlapping phase, three waves of activity coexist on the solar surface leading to the appearance and intensification of the odd zonal harmonic with $l = 5$. 
 
3. The maximum amplitude of the harmonic with $l=5$ was decreasing strongly during the past four cycles of solar activity, similar to the cycle amplitude obtained  from sunspots. The magnitude of the harmonic corresponding to the Gaussian coefficient $g_{50}$ is observed to fall especially strongly after 2000, which agrees with low Cycle 24. Of course, four cycles are hardly sufficient for convincing statistics, but it seems reasonable to expect that Cycle 25 will not be much higher than Cycle 24.

These conclusions were supported and refined by new observations. Such refinements, which appeared after the publication of [17], are given below and constitute one of the goals of this work.  

Another goal of this work is to form an idea of the activity cycles on stars of late spectral types using the solar cycle as an example.
Naturally, the observations of the magnetic activity on these stars are much less detailed than the scheme of the extended solar cycle described above. Therefore, it is important to understand to what extent this complex scheme can be compared with data on the stellar magnetic activity that are available now or can be obtained in the foreseeable future.

\section{The role and meaning of the fifth harmonic}

The central role in the proposed description of the phenomenon of the solar cycle belongs to the zonal harmonic $l=5$. Let's consider its behavior in more detail.

Figure 1 shows a synoptic map of the magnetic field. One can see the magnetic flux propagate from the latitudes $\theta= -30^\circ$   and $\theta= +30^\circ$ to the poles. This phenomenon, commonly referred to as Rush-to-the-Poles (RTTP), is mainly associated with large-scale fields. Upon reaching the poles, RTTP replaces the previous wave, and a reversal of the polar field occurs. Besides that, there is a well-known wave moving from mid latitudes to the equator. The polar and equatorial waves appear almost simultaneously, and have opposite predominant polarities of the magnetic field. At some moments, two Rush-to-the-Poles waves with opposite field signs coexist on the Sun. When one of them almost reaches the pole, the other only appears at mid latitudes. In such periods, three zones of alternating magnetic-field polarity exist in each hemisphere on the Sun.

\begin{figure}[!htbp]
{\includegraphics[width=\textwidth]{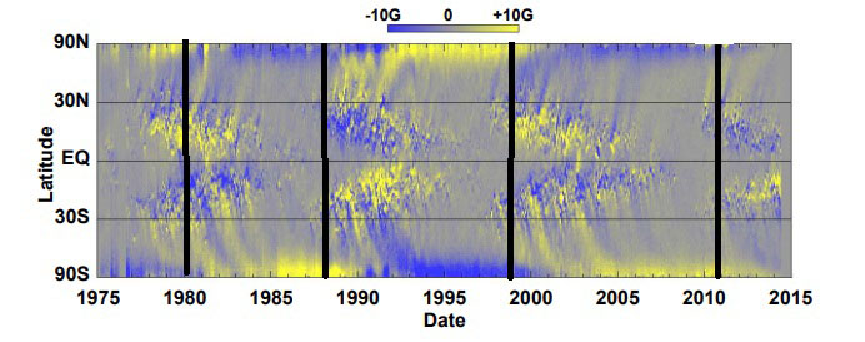}}
\caption{Supersynoptic map based on Kitt Pick Observatory data (positive polarity - white, negative polarity - dark). The vertical lines indicate the positions of the maxima of the fifth harmonic. The ordinate axis is the sine of latitude; the horizontal bars mark the latitudes of $30^{\circ}$, $0^{\circ}$, and $-30^{\circ}$.}
\label{fig:f1} 
\end{figure}

At first glance, it is not quite clear why we emphasize the role of one of the zonal harmonics. As an example, let's take a closer look at the supersynoptic map in the vicinity of 1999 (see Fig. 1). At the north pole, we can still see the wave of positive polarity, which formed in 1989 (i.e., in the growth phase of Cycle 22), and was moving to the pole during 13 years until 2002 (i.e., in fact, till the maximum of Cycle 23). At the same time, another wave of negative polarity existed at mid latitudes since 1999 and was also drifting poleward for 13 years until 2012 (i.e., until the ascending branch of cycle 24). In the same year (1999), the third wave appeared, which, unlike the first two waves, was not strictly unipolar, but nevertheless it had predominant positive polarity, corresponding to the polarity of the leading spots in active regions. This wave was propagating towards the equator. 

 As a result, we have three zones of alternating polarity in each hemisphere. As a whole, the interaction of these three magnetic systems continues from 1989 to 2005, i.e., 16 years. Thus, there is a period when the waves of all three types are present simultaneously. At this time, six intermittent zones are observed in general on the visible solar disk, which exactly corresponds to the zonal harmonic with $l=5$. We call this time interval the overlapping phase. We will see below that the overlapping phase can be quantitatively described by the fifth zonal harmonic, which, in this context, can be called the height of the overlapping zone.

Figure 1 was plotted using WSO (John Wilcox Stanford Observatory) data for the period from May 1976 (Carrington rotation 1641) up to the present (http://wso.stanford.edu/forms/prsyn.html). The observed solar surface magnetic field was expanded into harmonic coefficients, and the time evolution of the corresponding coefficients was represented over the last four cycles of solar activity. We analyzed WSO synoptic maps of the line-of-sight component of the photospheric magnetic field expanded into the series of associated Legendre polynomials $ P_{ml}$ [18].

All components of the magnetic field at a given point of a spherical layer between the photosphere and the so-called source surface can be reconstructed under the potential approximation from the line-of-sight component. The source surface is defined as a spherical surface, on which all magnetic lines become radial. It is assumed that it lies at a distance $R_s = 2.5 R_o$ from the center of the Sun ($R_0$ is the solar radius). In this case, the classical scheme was used, which assumes that the potential approximation can also be applied to the lower boundary surface, i.e., the photosphere level. Figure 2 shows the actual field structure of the fifth harmonic on June 1, 1999. All six zones are clearly visible.

\begin{figure}[!htbp]
{\includegraphics[width=\textwidth]{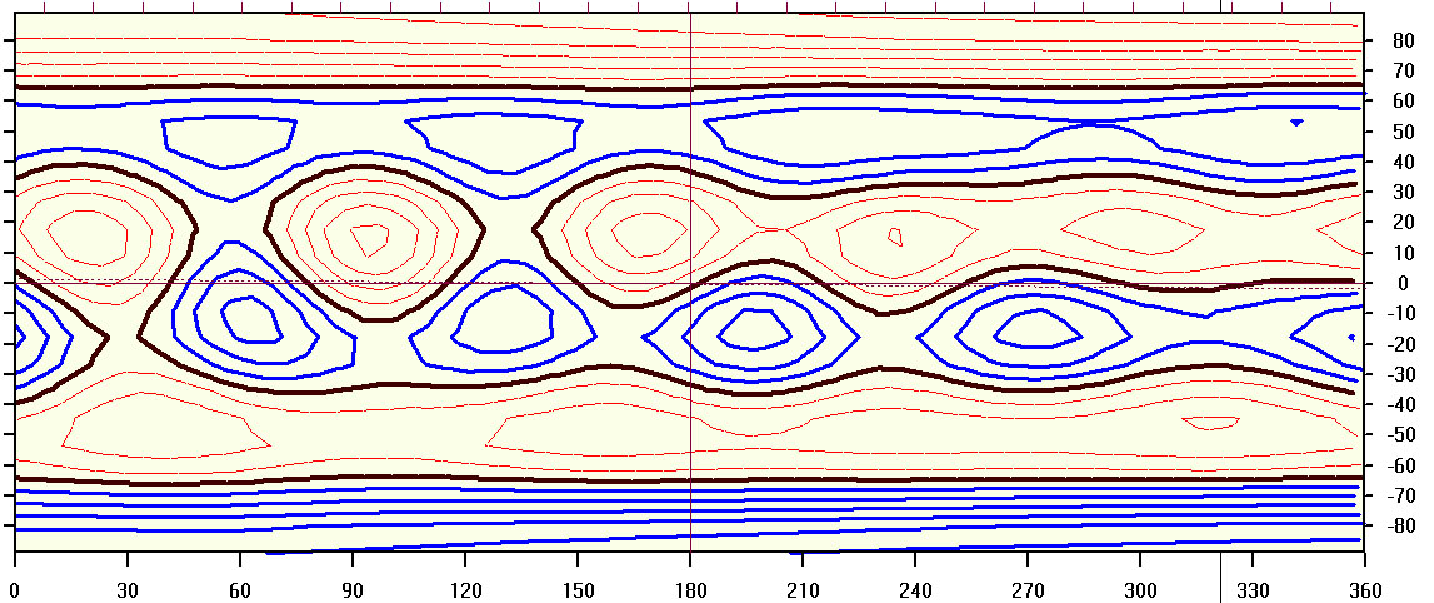}}
\caption{The field structure on the photosphere with the fifth harmonic alone taken into account. One rotation is shown, centered on June 1, 1999.}
\label{fig:f2_bw}
\end{figure}

Next, following [19] (see also [20]), we calculate the mean square radial component of the magnetic field on a sphere of radius $R$. Using the orthogonality of the Legendre polynomials, we calculate integrals over a spherical surface and obtain the result in closed form. Below, we give as an example the result of calculations on the photosphere  $i(B_r)|_{R_0}$ and on the source surface $i(B_r)|_{R_s}$:

\begin{equation}
i(B_r)|_{R_0}=\sum_{lm}\frac{(l+1+l\zeta^{2l+1})^2}{2l+1}(g^2_{lm}+h^2_{lm}),
\label{eq1}
\end{equation}

\begin{equation}
i(B_r)|_{R_s}=\sum_{lm}(2l+1)\zeta^{2l+4}(g^2_{lm}+h^2_{lm}),
\label{eq2}
\end{equation}

\noindent where $\zeta=R_0/R_s$. Hence, the contribution of the $l^{th}$ mode to the mean solar magnetic field includes all coefficients with  $l$. We assume $R_s=2.5 R_0$; i.e., $\zeta=0.4$. Choosing appropriate values of $l$ and $m$ in these relations, we obtain the contribution of the corresponding harmonics to the mean square magnetic field on the corresponding surface. The calculation results are shown in Figure 3.

\begin{figure}[!htbp]
{\includegraphics[width=\textwidth]{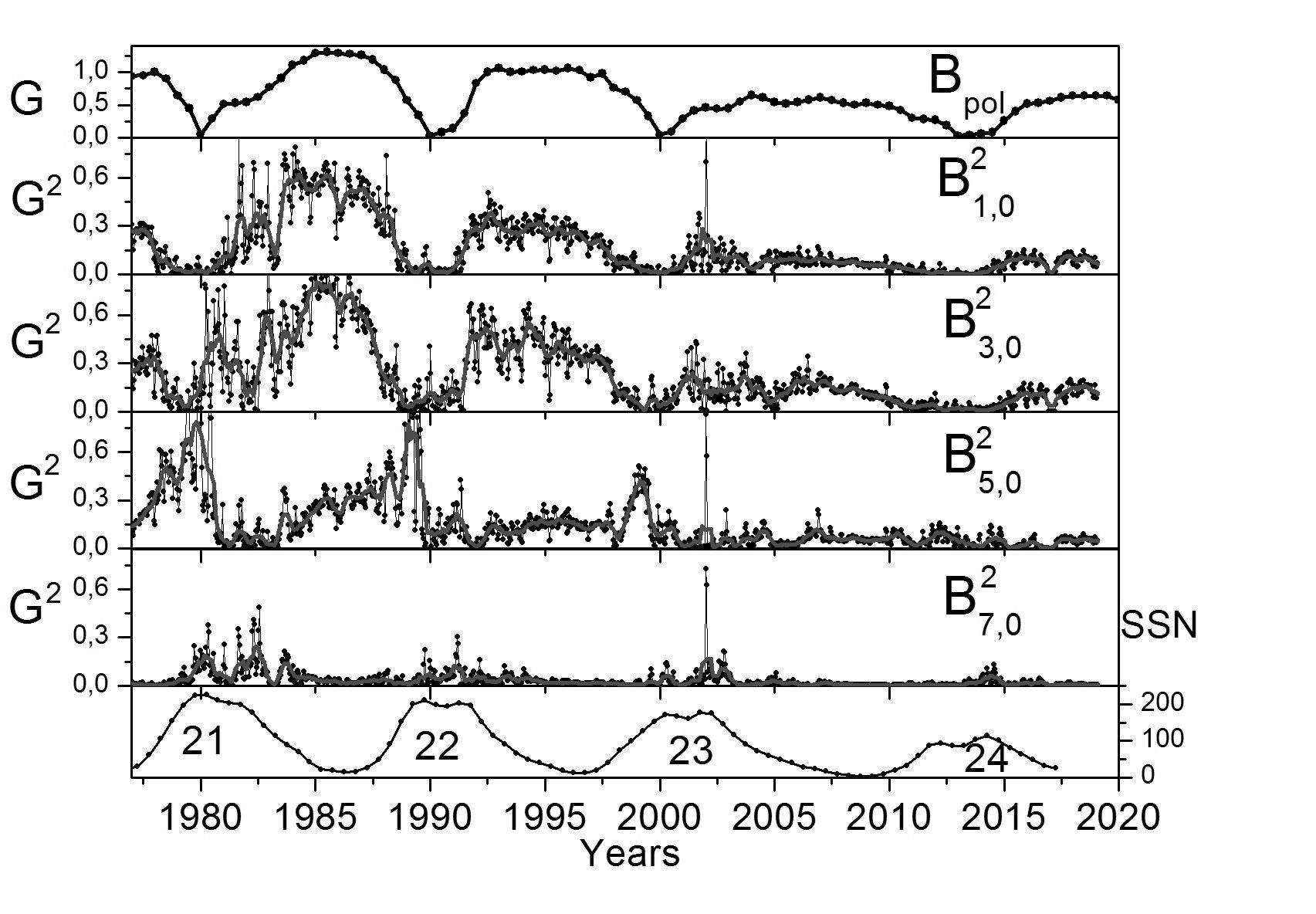}}
\caption{Time evolution of the mean square magnetic field associated with the first odd axisymmetric harmonics up to $l=7$ [21]. The lowest panel shows the time variation in sunspot numbers (SSN) during the solar cycle; the top panel represents the time evolution of the polar magnetic field.}
\label{fig:f3}
\end{figure}

The harmonics with $l=1$ and $l=3$ behave in accordance with the classical dynamo theory; i.e., they follow the evolution of the polar magnetic field and change in anti-phase with the sunspot numbers. On the contrary, the harmonic with $l=5$ behaves in essentially different way and is maximum just at the beginning of the minimum of the polar magnetic field [17].

To sum up, we can say that shortly before the sign reversal of the polar magnetic field, both the large-scale and the relatively small-scale magnetic fields appear almost simultaneously on the photosphere surface (the overlapping phase). Then, however, these fields behave differently. The large-scale component as a unipolar feature moves to the poles (Rush-to-the-Poles phenomenon). As a result, two waves of opposite polarities appear, propagating on the solar surface. On the contrary, the local magnetic field is represented by bipolar features propagating towards the equator. Formally, the separation of both types of waves is quantitatively related to the harmonic with $l=5$.

The maximum of the fifth harmonic is displayed as a rather sharp peak coinciding in time with a relatively short period of coexistence of all three waves in each hemisphere. 

At the time when the nearest wave reaches the pole and disappears, a reversal of the polar field occurs. From that moment on, only two waves are observed on the Sun in each hemisphere, and the harmonic with $l=5$ virtually disappears.

Note that the maxima of the zonal harmonic with $l=5$ are clearly defined in Cycles 21, 22, and 23, but are almost invisible in Cycle 24, when we are dealing with an extended minimum and a rather indistinct overall structure after 2012 (see Fig.1). One can see that    for the fifth harmonic, $B^2$ does not exceed 0.1 $G^2$ during the entire period after 2000. It should be noted that Cycle 24 was anomalous in many respects. In particular, it was strongly distorted by asymmetry. The maxima in the northern and southern hemispheres were so much separated in time that the total number of sunspots was much lower than in other cycles.

The behavior of the fifth harmonic until 2022 strongly resembled the situation before Сycle 24; therefore, we suggested in [17] that Cycle 25 cannot be significantly higher than Cycle 24.

\section{Estimates of the height and date of Cycle 25}

The prospects for Cycle 25 have become somewhat more definite lately, after our work [17] was already published. First of all, we can use data on the behaviour of the polar field. The connection between the intensity of the polar field and the height of the following solar cycle is best substantiated physically and relies on the basic concepts of the solar dynamo (for details, see extensive reviews [22, 23].

The linear relationship between the maximum value of the polar field $B_{\rm pol}$ at the end of the previous cycle (usually near the SSN minimum) and the height of the following sunspot cycle based on data for the last four cycles is shown in Fig. 4 and is approximated by the relation 

\begin{equation}
SSN_{max}= 36.405+1.3666 B_{\rm pol}.
\label{eq3}
\end{equation}
The observed maximum value of $B_{\rm pol}$ was reached in the summer of 2019 and was equal to 65  $\mu T$ (http://wso.stanford.edu/Polar.html). Hence, the predicted value of SSN in Cycle 25 is 125.2. This is only 8.8 units higher than in Cycle 24 (116.4).

\begin{figure}[!htbp]
{\includegraphics[width=\textwidth]{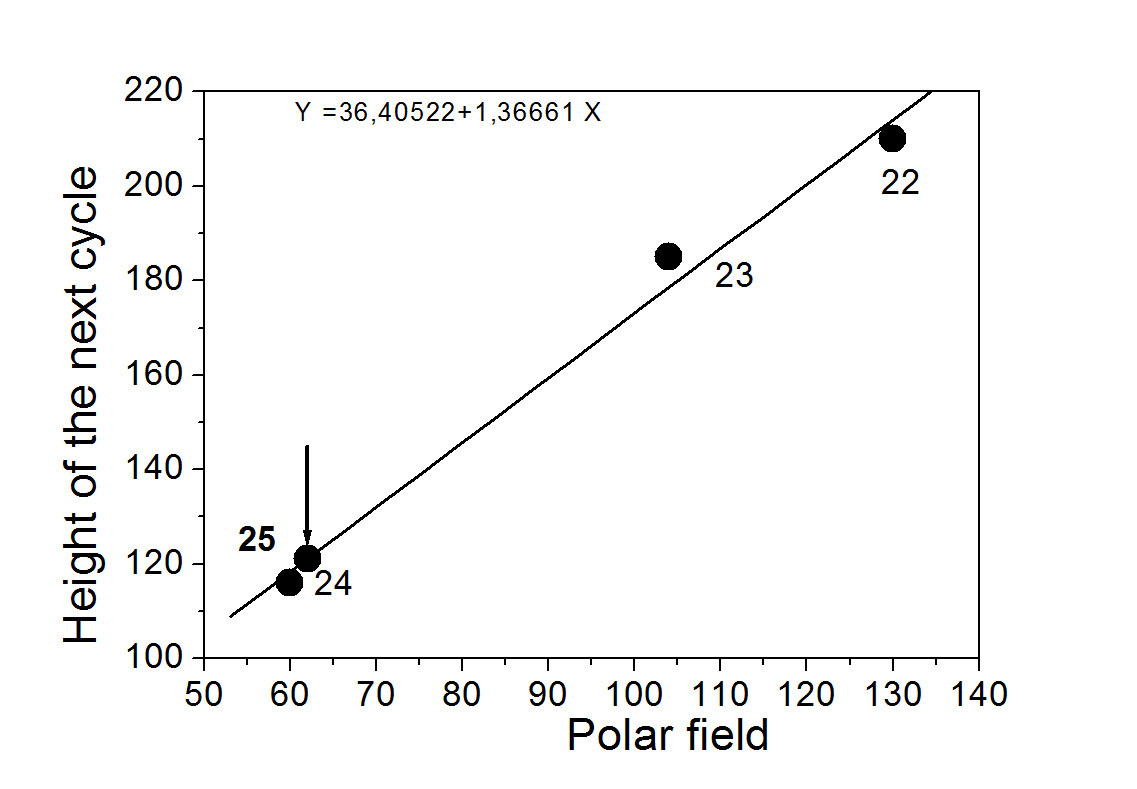}}
\caption{Relationship between the maximum value of the polar field, $B_{\rm pol}$, before the cycle and the height of the coming sunspot cycle. The expected value for Cycle 25 is indicated by the arrow.}
\label{fig:f4}
\end{figure}

\begin{figure}[!thbp]
{\includegraphics[width=\textwidth]{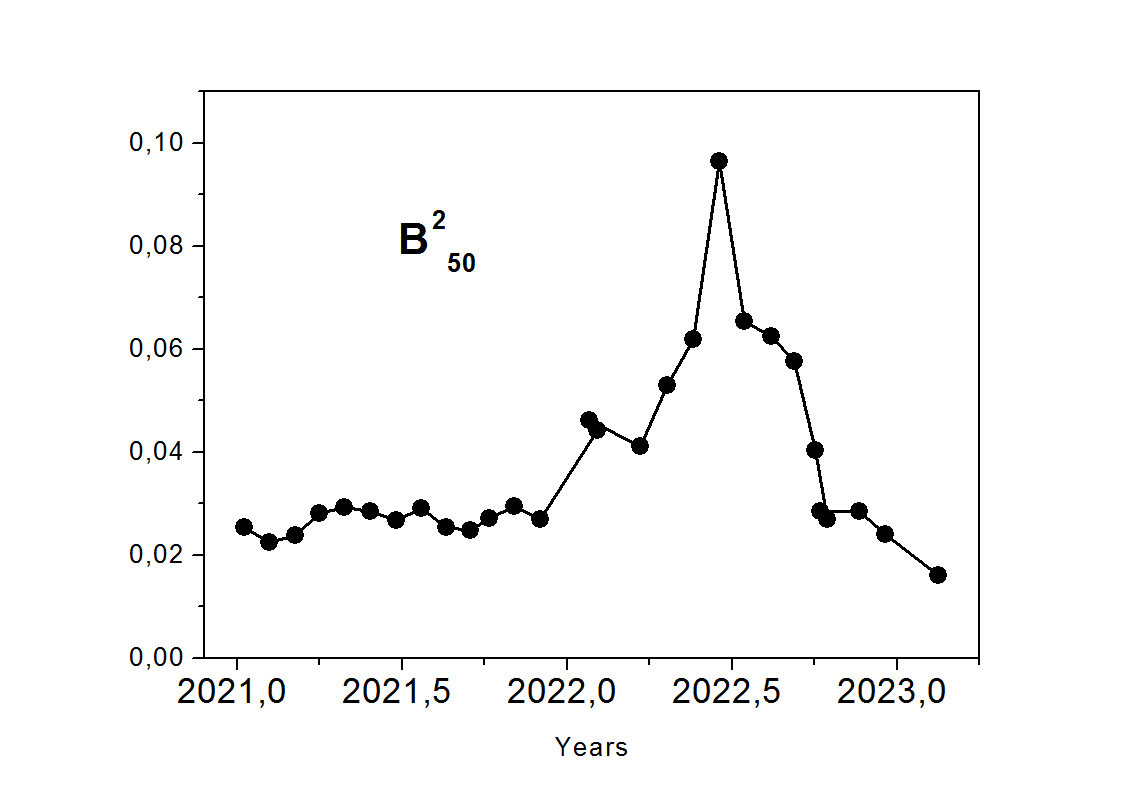}}
\caption{Variations in the squared amplitude of the fifth zonal harmonic over the past two years.}
\label{fig:f5}
\end{figure}
It is to be noted that the variation of the fifth harmonic also became more regular (see Fig. 5). At the same time, its maximum value reached in 2022.5 was 0.98 $G^2$, which is almost an order of magnitude lower than the maximum values of this harmonic in Cycles 21 and 22. This probably  indicates that the mechanism of generation of the three waves mentioned above does not work efficiently enough.

\begin{figure}[!htbp]
{\includegraphics[width=\textwidth]{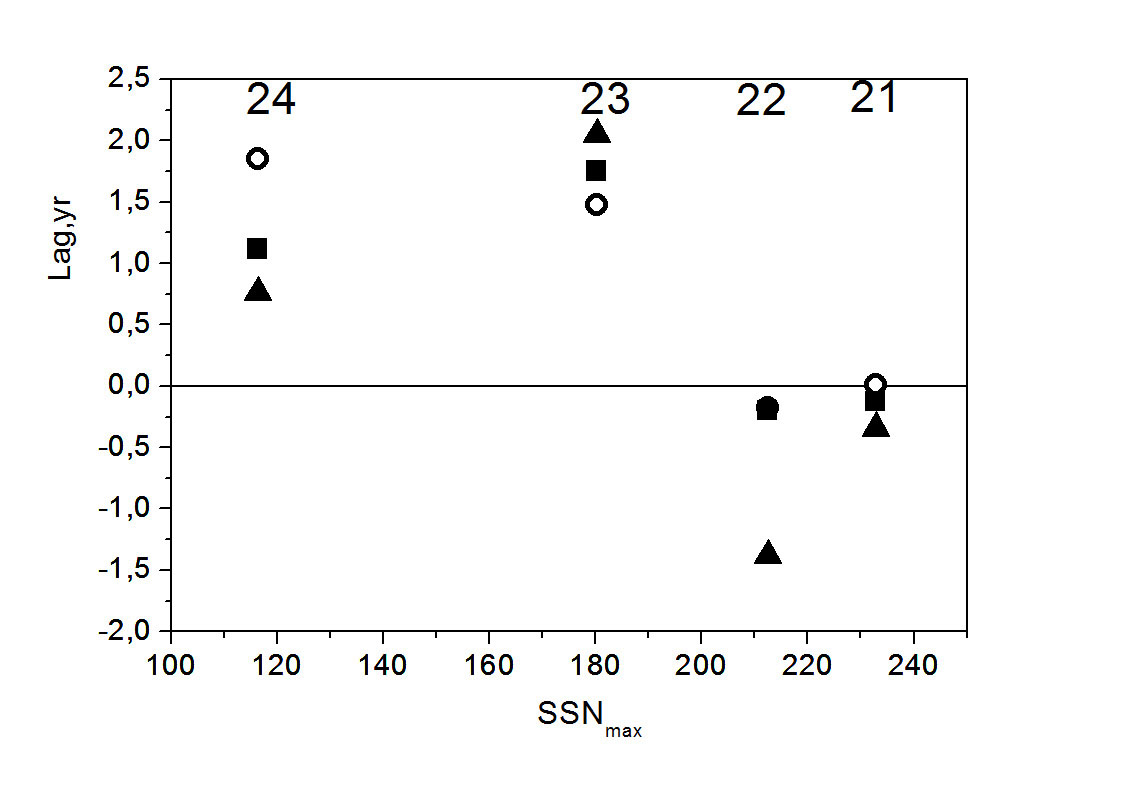}}
\caption{The lag (difference between  $T_{\rm max}$  and $T_P$,): the positive values indicate that the polarity reversal occurred prior to the maximum of the sunspot cycle; the negative values show that the polarity reversal was delayed. The lag values were calculated separately for the north (open circles) and the south pole (triangles). The average reversal dates (avg) are shown with squares.}
\label{fig:f6}
\end{figure}

As the dates of the maximum, we can use the dates of reversal of the polar field. According to the classical model, the dates of the polarity reversals and cycle maxima must coincide or at least be close in time. Figure 6 illustrates the time delay between the polarity reversals $T_P$ and the $SSN(T_{\rm max})$ dates. The lag was calculated as the difference between $T_{\rm max}$ and $T_P$. The positive values indicate that the polarity reversal occurred prior to the maximum of the sunspot cycle; the negative values show that the polarity reversal was delayed. The lag values were calculated separately for the north (open circles) and the south pole (triangles). The average reversal dates (avg) are shown with squares.

It turned out that in the relatively high (or at least, medium) Cycles 21 and 22, the SSN maximum  either coincided with or took place slightly ahead of the polarity reversal, while in the low Cycles 23 and, especially, 24, the polarity reversal occurred 1.0 - 1.5 years earlier. Note that the very concept of the exact date of the cycle maximum is rather conventional. Usually, the date of the maximum is the month when the curve of the sunspot numbers smoothed over 13 months reaches its maximum value.

However, there is a long time interval when the values on the curve differ from the maximum value rather weakly, by only a few units. Therefore, it would be more correct to speak of a maximum phase. Now that we already know the polarity reversal date at the beginning of Cycle 25 in the north hemisphere (March 2023), we can expect the cycle maximum between the end of 2023 and the second half of 2024, which is much earlier than predicted from a simple 11-year periodicity (2025 or early 2026).

Thus, the maximum may come earlier than predicted (already in 2023), and its height is expected to be 125.2, i.e., only 9 units more than in Cycle 24.  The earliest date of the onset of the maximum phase is November-December 2023.
 
Another question the future observations may answer is whether the Gnevyshev-Ohl rule remains valid for this cycle. The point is that there are several formulations of this rule [24]. The original rule states that the sum of sunspot numbers per cycle in even cycles is less than in subsequent odd cycles [25]. Later, it was extended to the monthly maximum sunspot numbers per cycle. The rule in the latter formulation was broken in the pair of Cycles 22-23.

If our calculations are correct, the Gnevyshev-Ohl rule in this formulation will hold in the pair of Cycles 24-25, although Cycle 25 will only slightly exceed Cycle 24 in height. However, it follows from our calculations that the growth branch in Cycle 25 will be very short (4.0–4.5 years), i.e., much shorter than the growth branch of Cycle 24 (about 5.5 years). Therefore, there is no certainty that the rule for the integral sum of sunspot numbers per cycle in its classical form will hold in the pair of Cycles 24-25. Here, a very long maximum phase in Cycle 24 (almost 3 years) can play a special role.

The results of this section are briefly summarized as express information in [26]. 

\section{A possible scenario for observing the structure of activity on late-type stars}

The structure of the solar cycle described above can be used not only to forecast the periodic solar activity. It can also indicate what exactly should be observed to better understand the activity on stars more or less similar to the Sun, i.e., the late-type stars. We know from spectrophotometric data and, first of all, from the data of the H-K Project [27] and conceptually similar projects [28] that a number of late-type stars exhibit activity cycles similar to the solar ones, in particular, their duration is comparable to the length of the solar cycle. Although the periodicity of solar type on the stars with similar physical characteristics seems natural, the discovery of these cycles was a fundamental achievement of astronomy. Note that on many stars observed within the framework of the mentioned projects, the detected time variations of the studied spectral lines do not have a pronounced periodicity. On some stars, secondary cycles were detected along with the main ones (see, for example, [29, 30]). The analysis of archival time series available for individual stars (e.g., [31, 32]) shows that instead of a cyclic behavior, fluctuations can form a continuous spectrum of variations.

It is believed that the stellar activity cycles and other similar variations are associated with inductive effects and rank among the phenomena studied in the framework of the hydromagnetic dynamo theory (although, there are stars with a relic magnetic field, as, apparently, are Ap stars (e.g., [33])). The variable activity of late-type stars is usually discussed within the paradigm that the magnetic field excited by the stellar dynamo has a configuration similar to that of the solar magnetic field. This point of view was characteristic of the first (successful) attempts to detect stellar cycles. In a few cases, the time-latitude pattern of the cyclic activity of particular stars could be reconstructed (at least roughly) from the available data, and it really turned out to be similar to the solar activity (see, for example, [34]). However, there are many stars of late spectral types whose magnetic-field structure and cyclic activity may differ very much  from those of the Sun. For example, it is widely believed that the magnetic activity of M-type dwarfs differs markedly from that of the Sun (e.g., [35]), although the question of what exactly is this difference remains open. Another example is the discovery of stars comparable to the Sun, but having kilogaussian magnetic fields (e.g., [36]) or producing much stronger flares than the Sun [37]. All said above makes us believe that the next step in the study of the stellar magnetism should comprise observations that will either show the possibility of excitation of essentially different magnetic configurations on stars or prove that all really existing configurations are similar to those of the Sun.

The existing theoretical models of the stellar dynamo allow various magnetic field configurations excited by dynamo mechanisms similar to that acting on the Sun. Virtually all scientists involved in modeling spherical dynamos have noted this under various circumstances (see [38] for one of the first references), although, of course, it is difficult to prove by numerical simulations that such configurations actually occur on stars. On the other hand, theoretical arguments can be provided in favor of the statement that all stars close to the Sun in their physical properties must have a magnetic field similar to that of the Sun (cf. [39]). Obviously, the described problem should be solved by observational methods, i.e., by constructing latitude-time diagrams of stellar activity based on a representative sample of observational data for stars of late spectral types. Since the problem is rather general, this sample does not have to contain only stars that are very similar in their properties to the Sun.
 
Fortunately, the available observational capabilities, first of all, the inverse Doppler imaging known since the 80s of the past century [40], allow the solution of such problems, especially taking into account that we are only interested in the latitudinal distribution of star spots, and there is no need to determine their sidereal longitude.

The proposed program involves long-term monitoring of stellar activity comparable in duration to the the H-K Project. (It is remarkable that this Project had started before the inverse Doppler imaging technique was developed). Such a program requires that we clearly define the particular features in the described scenario of a solar cycle that are to be compared with the stellar activity data, since the discovery of their entire variety is apparently the matter of a distant future. This is precisely the purpose of the present section.

The idea of the proposed monitoring is that, relying only on temperature data and restoring the latitude-time distribution of star spots over two or three cycles of stellar activity, we can to establish more or less confidently the similarity or dissimilarity between the solar activity and  the activity of a given star. This idea, is based on the fundamental zonal model of star spot distribution  developed by Alekseev and Gershberg [41]. But our goal is not to describe the distribution of spots using a limited number of parameters, but to find out what magnetic configurations they are associated with and what can this tell us about the operation of the stellar dynamo.

Instead of reconstructing the positions of individual star spots on the basis of the reverse Doppler images, it seems enough to restore the position of the latitude band, $\Theta$, in which these spots arise, its half-width $\theta$, its contrast with the surrounding surface, $K$, and the evolution of these parameters over a stellar activity cycle. It is also necessary to find out whether such a band is observed separately in each hemisphere or whether a single band of activity common to both hemispheres appears in the vicinity of the stellar equator.

Let us explain how the behavior of these parameters helps us understand whether the cyclic activity of a given star resembles that of the Sun. The presence of a single activity band near the stellar equator, which does not split into two individual bands in each hemisphere, would indicate that we are dealing with a magnetic configuration with quadrupole rather than dipole symmetry. This test is based on the fact that the toroidal component of the mean magnetic field with dipole symmetry vanishes at the stellar equator, while in the case of the quadrupole symmetry it need not vanish. Similar configurations appear in stellar dynamo models along with ordinary solar-type configurations, but have not yet been detected in observational practice. They are discussed in literature (e.g., [42]). In archival observations of solar activity, a similar episode was reported in the 18th century [43], although the quality of observations of that time does not allow us to confidently insist on its reality.

Estimating the half-width of the magnetic activity band, $\theta$, we can determine what zonal harmonic of the magnetic field is excited by the stellar dynamo mechanism. For example, as shown by observations of the mean magnetic field on the solar surface, the harmonic $l=5$ dominates the Sun, which corresponds to the width of the activity band of about $30^\circ$. Indeed, a zonal harmonic with index $l$ divides the sphere into $l+1$ bands. The dipole has two bands (two hemispheres), the quadrupole has three bands, etc. For $l=5$, we have six bands of width $180^\circ /6 =30^\circ$ (Fig. 2) (cf. [ 17]). This is  the generally recognized zone of local activity. The Carrington rotation rate corresponds exactly to the middle of this zone - $16^\circ$. The sunspot cycle usually begins approximately at the upper boundary of this zone — $30^\circ$.

Changes in the position of the central latitude of the activity band, $\Theta (t)$, can help us understand whether the activity waves propagate towards the stellar equator, as happens on the Sun, or towards the poles of a star, which is also quite possible within the concept of the stellar dynamo at a certain distribution its sources (see, for example, [44]).

And finally, the contrast of the activity band against the surrounding stellar surface can show how much the flare activity of a star may differ from that of the Sun (cf. [45]).

It is natural to reconstruct the magnetic configuration on stars from observations of light curves in different spectral ranges on the basis of the reverse Doppler imaging technique. In other words, instead of choosing the initial set of the spot formation zones with the parameters corresponding to the observed light curves, it is proposed to consider the probability density of the spot formation at a given latitude (or a similar quantity) as a continuous function of latitude. This function is found from the solution of the corresponding inverse problem. In the process, we propose that the residual minimization procedure applied to determine the discrepancy between the model and observations include terms that depend on the number and size of the desired distribution by using the Tikhonov regularization or maximum entropy methods. Experience shows that the use of such methods for mapping the temperature distribution on the stellar surface allows us to avoid uneven detalization of the obtained pattern generated by observational errors rather than real observations.

Of course, we describe a minimal observation program. One can supplement temperature observations with magnetic data, i.e. support the inverse Doppler images with inverse Zeeman-Doppler images. This makes it possible to significantly expand the capabilities of monitoring, e.g., to distinguish between the quadrupole and dipole magnetic fields much more confidently, since the polarity rule  similar to the Hale's one sounds different  for these configurations. For quadrupole configurations, the dominant polarity in both hemispheres of the star coincides, and the polarity reversal occurs at the transition from one cycle to the other. Recall that,  originally, the Hale polarity rule was formulated shortly after the first successful measurements of the sunspot magnetic field, without years of preliminary observations.

A representative sample for such a study can be formed in different ways. There is no need to focus only on solar-type stars; however,
it is important that these stars have convective zones and exhibit noticeable differential rotation, since otherwise  the question of a solar-type dynamo becomes meaningless. Implementation of such a project is a very difficult task. The difficulty lies not so much in developing new expensive equipment, but, as seen from the experience of the H-K Project, in recruiting a small but well-motivated team of researchers and ensuring a succession of scientific generations in the team.

Let us briefly formulate the difference between the proposed observation program and the programs currently being developed. 

1. We propose to focus on a long-term monitoring of temperature variations on a large but limited number of selected stars similar to the Sun in their physical properties rather than expanding the list of stars under investigation and focusing on the variety of methods available to characterize the magnetic field of each star at a given moment in time.

2. Shift the main attention from studying the field distribution over the stellar surface at a given time to studying the latitude-time distribution of the magnetic field over time intervals comparable to several cycles of stellar activity.

\section*{Conclusions}

Let us summarize the potential capabilities of the updated description of the solar cycle proposed in the recent works of our team and discussed above. 

Firstly, this description may allow a more accurate and reliable prediction of the evolution of the cyclic activity of the Sun. Such a forecast may be tested in the near future. 

Secondly, this description can serve as the basis of a program for studying the magnetic activity of stars more or less similar to the Sun in their physical characteristics. Despite the significant effort required to implement such a program (similar to the H-K Project), it seems feasible within a limited time frame and will allow us to radically improve our knowledge of the magnetic activity of stars.

\section*{Acknowledgements}

The authors are grateful to Dr. T.Hoeksema for open access to data on the site http://wso.stanford.edu. We also express our gratitude to the Ministry of Science and Higher Education for support in the frames of grants 075-15-2020-780 (VNO and MMK) and 075-15-2022-284 (DDS).

$$$$

{\bf References}

$$$$

1. R.C.Altrock, Bulletin of the American Astronomical Society, 20, 723 (1988).

2. P.R.Wilson, R.C.Altrock, K.L.Harvey, S.F.Martin, H.B.Snodgrass, Nature, 333 (6175), 748-750 (1988).

3. J.-L.Leroy, J.-C.Noens,  Astron. Astrophys., 120 (2), L1-2 (1983).

4. A.Kosoviche, V.Pipin, A.Getling, in {\it American Astronomical Society Meeting Abstracts}, 2021, p. 304.

5. S.W.McIntosh et al.,  Sol. Phys., 296, 189 (2021).

6. G.Bocchino, Mem. Soc. Astron. Italiana, 6, 479 (1933).

7. R.Hansen, S.Hansen, Sol. Phys., 44, 225 (1975). 

8. P.S.McIntosh, in Astronomical Society of the Pacific Conference Series, ed. by K.L.Harvey, {\it The Solar Cycle}, 27, 14 (1992). 

9. A.G.Tlatov, K.M.Kuzanyan, V.V.Vasil’yeva, Sol. Phys., 291, 1115 (2016).

10. K.L.Harvey, S.F.Martin,  Sol. Phys., 32, 389 (1973).

11. R.C.Altrock,  Sol. Phys., 170, 411 (1997).

12. R.Howard, B.J.Labonte, ApJ, 239, L33 (1980).

13. H.B.Snodgrass, P.R.Wilson, Nature, 328, 696 (1987).

14. P.R.Wilson, {\it Solar and stellar activity cycles}, Cambridge Astrophysics Series, 24 (1994).

15. P.N.Mayaud,  J. Geophys. Res., 80, 111 (1975).

16. J.P.Legrand, P.A.Simon, Sol. Phys., 70, 173 (1981).

17. V.N.Obridko, A.S.Shibalova, D.D.Sokoloff, MNRAS, 523 (1), 982-990 (2023).

18. V.N.Obridko et al., MNRAS, 492 (4), 5582-5591 (2020).

19. V.N.Obridko, F.A.Yermakov,  Astron. Tsir., 1539, 24  (1989). 

20. V.N.Obridko, B.D.Shelting, Sol. Phys., 137, 167 (1992).

21. V.N.Obridko et al., MNRAS, 504 (4), 4990-5000 (2021).

22. D.Nandy et al., Progress in Earth and Planetary Science, 8 (1), article id.40 (2021).

23. D.Nandy, Sol. Phys., 296(3), article id.54 (2021).

24. Yu.A.Nagovitsyn, E.Yu.Nagovitsyna, V.V.Makarova,  Pisma v AZh, 35 (8), 625 (2009).

25. M.N.Gnevyshev, A.I.Ohl, Aston. Zh., 23 (1), 18 (1948).

26. V.Obridko, D.Sokoloff, M.Katsova, Astron. Tsirk., 1658, 4 pp. (2023).

27. S.L.Baliunas et al., ApJ, 438, 269 (1995).

28. A.C.Baum et al., AJ, 163 (4), 9 pp. (2022), article id. 183.

29. E.Bohm-Vitense, ApJ, 657, 486 (2007).

30. A.Suarez Mascareno, R.Rebolo, J.I.Gonzalez Hernandez, Astron. Astrophys., 595, A12 (2016).

31. R.A.Stepanov et al., MNRAS, 495 (4), 3788-3794 (2020).

32. N.I.Bondar’., M.M.Katsova, Geomagn. Aeron., 62 (7), 919-922 (2022).

33. G.Mathys, I.I.Romanyuk, D.O.Kudryavtsev et al., Astron. Astrophys., 586A, 85 (2016).

34. M.M.Katsova, M.A.Livshits, W.Soon, S.L.Baliunas, D.D.Sokoloff, New Astronomy, 15 (2), 274-281 (2010). 

35. D.Shulyak, D.Sokoloff, L.Kitchatinov, D.Moss, MNRAS, 449 (4), 3471-3478 (2015).

36. O.Kochukhov, Astron. Astrophys. Rev., 29 (1), (2021), article id.1.

37. H.Maehara et al., Earth, Planets and Space, 67, pp. (2015), id.125006.

38. R.L.Jennings, N.O.Weiss, MNRAS, 252, 249-260 (1991).

39. L.L.Kitchatinov, Research in Astronomy and Astrophysics, 22 (12), 13 pp. (2022), id.125006.

40. A.V.Goncharsky, Astron. Zh., 59, 1146-1156 (1982).

41. I.Yu.Alekseev, R.E.Gershberg, Astron. Zh., 73 (4), 589-597 (1996).

42. D.Moss, S.H.Saar, D.Sokoloff, MNRAS, 388 (1), 416-420 (2008).

43. D.Sokoloff et al., in {\it Solar and Stellar Variability: Impact on Earth and Planets}, (Proc. of the International Astronomical Union, IAU Symposium), 264, 111-119 (2010).

44. E.Maiewski, H.Malova, V.Popov, D.Sokoloff, E.Yushkov, Sol.Phys., 297, 11, (2022), article id.150.

45. M.M.Katsova, V.N.Obridko, D.D.Sokoloff, I.M.Livshits, ApJ, 936, 1, 9 pp. (2022), article id. 49.

\end{document}